\documentclass[aps,pra,reprint,showpacs,nofootinbib,superscriptaddress,longbibliography]{revtex4-2}
\usepackage{graphicx,comment}
\usepackage{amsfonts}
\usepackage{amsmath,amssymb}
\usepackage{bm}
\usepackage{xurl}
\usepackage{qcircuit}
\usepackage{braket}
\usepackage{color}
\usepackage{epstopdf}
\usepackage{subfig}
\usepackage{epsfig}
\usepackage{float}
\usepackage{verbatim}
\usepackage{hyperref}
\usepackage{multirow}
\usepackage[table]{xcolor}
\usepackage{wrapfig}
\usepackage[normalem]{ulem}
\hypersetup{
     colorlinks   = true,
     citecolor    = blue
}
\usepackage{lineno}
\usepackage[thicklines]{cancel}
\usepackage{color, colortbl}
\definecolor{LightCyan}{rgb}{0.88,1,1}
\usepackage{url}
\usepackage{dcolumn}
\usepackage[utf8]{inputenc} 
\usepackage[T1]{fontenc}
{\rm }

\def\be{\begin{equation}}
 \def\ee{\end{equation}}
 \def\bea{\begin{eqnarray}}
 \def\eea{\end{eqnarray}}
\def\2{\frac{1}{2}}

\def\4{\frac{1}{4}}

\begin{document}

\title{Solving Power Grid Optimization Problems with Rydberg Atoms}

\author{Nora Bauer}
\email{nbauer1@vols.utk.edu}
\affiliation{Department of Physics and Astronomy,  The University of Tennessee, Knoxville, TN 37996-1200, USA}

\author{K\"ubra Yeter-Aydeniz}
\email{kyeteraydeniz@mitre.org}
\affiliation{Emerging Technologies and Physical Sciences Department, The MITRE Corporation, 7515 Colshire Drive, McLean, Virginia 22102-7539, USA}

\author{Elias Kokkas}
\email{ikokkas@vols.utk.edu}
\affiliation{Department of Physics and Astronomy,  The University of Tennessee, Knoxville, TN 37996-1200, USA}
\author{George Siopsis}
\email{siopsis@tennessee.edu}
\affiliation{Department of Physics and Astronomy,  The University of Tennessee, Knoxville, TN 37996-1200, USA}

\date{\today}

\begin{abstract}
The rapid development of neutral atom quantum  hardware provides a unique opportunity to design hardware-centered algorithms for solving real-world problems aimed at establishing quantum utility. In this work, we study the performance of two such algorithms on solving MaxCut problem for various weighted graphs.
% finding maximum power sections of an electric grid
% power grid and electric vehicle charging optimization problem. 
The first method uses a state-of-the-art machine learning tool to optimize the pulse shape and embedding of the graph using an adiabatic Ansatz to find the ground state.  We tested the performance of this method on finding maximum power section task of the IEEE 9-bus power system and obtaining MaxCut of randomly generated problems of size up to 12 on the Aquila quantum processor. To the best of our knowledge, this work presents the first MaxCut results on Quera's Aquila quantum hardware. Our experiments run on Aquila demonstrate that even though the probability of obtaining the solution is reduced, one can still solve the MaxCut problem on cloud-accessed neutral atom quantum hardware. The second method uses local detuning, which is an emergent update on the Aquila hardware, to obtain a near exact realization of the standard QAOA Ansatz with similar performance. Finally, we study the fidelity throughout the time evolution realized in the adiabatic method as a benchmark for the IEEE 9-bus power grid graph state. 
% \textcolor{red}{Write an inspiring abstract.} We solve combinatorial optimization problems using Rydberg Atoms and apply them to power grids and electric vehicle (EV) charging stations.

\end{abstract}

\maketitle

%\onecolumngrid
\section{Introduction}

Neutral atom quantum computers are a promising candidate for quantum optimization and simulation, offering large numbers of qubits, flexible connectivity, and longer coherence times. Current hardware, such as QuEra's Aquila processor \cite{wurtz2023aquila}, offer analog control, where instead of implementing circuits comprised of discrete gates as in digital quantum computers, the user programs a time-dependent Hamiltonian to evolve the quantum system. The Aquila processor offers 256 Rb-87 atoms as physical qubits that can be configured in a 2-dimensional array, and their evolution is governed by Rabi frequency and detuning pulses, which can be programmed as time-dependent functions. The atoms experience a Rydberg interaction between excited states that depends on the physical distance between them. 

Neutral atom quantum computers are also capable of acting as logical processors with quantum error correction. In Ref.\ \cite{Bluvstein2024}, a neutral atom quantum system with 280 physical qubits was used to encode up to 48 logical qubits and 40 color code qubits. However, this procedure involved universal digital single-qubit gates, arbitrary connectivity, and mid-circuit readout, which are not publicly available to users on current hardware. In the near term, the constraint of a fixed Hamiltonian as well as the flexibility of atom placement offer a potential for efficient solutions to certain problems in quantum simulations and optimization. The Aquila processor can be used to engineer quantum states such as Bell states, cat and Greenberger–Horne–Zeilinger (GHZ) states \cite{omran2019generation,q-ctrl}, as well as antiferromagnetic spin states with domain wall defects \cite{balewski2024engineering,q-ctrl-2}. The Aquila processor has been used to prepare and study a two-dimensional quantum spin liquid with $\mathbb{Z}_2$ topological order and anyon condensation \cite{Lukin2022}. Additionally, neutral atom quantum computers have a native realization of the maximal independent set (MIS) combinatorial optimization problem \cite{ebadi2022quantum,Kim2022}, where the ground state solution of the quantum Hamiltonian can be realized by optimizing an adiabatic evolution. The capabilities can be expanded to non-native independent set problems which can also be solved by optimizing a non-adiabatic pulse sequence \cite{balewski2024engineering}. Recent work using an adiabatic approach utilized a graph embedding optimization procedure and machine learning model to train the pulse parameters for solving combinatorial graph problems on a neutral atom quantum computer \cite{coelho2022efficient}. They also measured the Q-score benchmark value which quantifies the performances of a given device or method in solving a specific combinatorial optimization problem, such as the Maximum Cut (MaxCut) problem. In neutral atom quantum computers, state preparation is usually achieved by pulse shaping, i.e., optimizing the time-dependence of the Hamiltonian. This process can be extremely costly, as it requires a lot of sampling of the final state in the quantum processor (or in its emulator). Recent developments in software designed for analog computers expand capacities for differentiable circuit optimization with applications in solving differential equations and optimization problems \cite{seitz2024qadence}. 

Many real-world optimization problems can be mapped to weighted MaxCut problems. Power grid optimization \cite{jing2023data} and EV charging station placement problems \cite{jing2022dynamics,jing2022quantum} are of particular interest for quantum utility, and quadratic unconstrained binary optimization (QUBO) formulations have been solved using the quantum approximate optimization algorithm (QAOA) on digital hardware. Weighted MaxCut problems are particularly suitable for neutral atom hardware because the atom placements can be adjusted to encode weights efficiently. 

In order to compare across platforms, such as superconducting and ion trap quantum hardware, it is necessary to develop and test benchmarks on neutral atom quantum hardware. The Quantum Economic Development Consortium (QED-C) algorithmic benchmark set has been used to quantify the performance of simulated noisy neutral atoms on problems such as the quantum Fourier transform (QFT), Deutsch-Jozsa, and phase estimation algorithms \cite{wagner2023benchmarking,abbas2023quantum}. There  has also been a benchmark developed for the generation of GHZ states on a neutral atom quantum computer \cite{graham2022multi,Shaw2024}. However, these studies use digital quantum gates rather than analog quantum computing. Benchmarks developed for analog computing \cite{shaw2023benchmarking,mark2023benchmarking} involve measuring the fidelity of the experimental system compared to the noiseless system as a function of time throughout the evolution of the system. 

In this work, we studied the performance of two different methods for solving weighted MaxCut problems corresponding to the power grid and EV charging optimization problems. The first method used pulse shaping of an adiabatic Ansatz via machine learning and a graph embedding algorithm that optimized the atom placements based on graph structures. We studied problem sizes up to 13 graph nodes and find over 95\% probability of measuring the ground state solution on the graphs tested, in addition to reasonable scaling with the number of steps required to find the solution. We performed numerical simulations and also tested the performance on the Aquila hardware. The second method used QAOA with local detuning control on atoms, which is a near-term update on the Aquila hardware. Using this method, the detuning is not a free parameter anymore but depends on the geometry of the graph. By doing so, we could eliminate all the linear $Z$ terms from the Hamiltonian of the neutral atoms, which gave a native realization of the weighted MaxCut cost function. We performed the QAOA for random graphs placed on square and honeycomb lattices up to 13 vertices, and solved the MaxCut problem with probability 83\% in the best case scenario. We also performed a fidelity benchmarking test on the power grid optimization problem on the hardware. 

Our discussion is organized as follows. In Section \ref{sec:usecase}, we introduce the power grid and EV charging optimization problems and their mapping to the weighted MaxCut problem. In Section \ref{sec:2methods}, we discuss two methods used to solve the optimization problems on analogue computers and provide experimental and numerical results. Section \ref{sec:benchmarking} focuses on the neutral atom benchmarking procedure where we provide experimental results solving the optimization problems on the Aquila hardware. Finally in Section \ref{sec:conclusion}, we summarize our results and provide future directions of research. 

\section{Use Cases}
\label{sec:usecase}

Electric power systems are critical in industrial societies. Operation of large transmission and distribution systems requires resource intensive computational processes. Optimization is one of these computational processes which plays an essential role in power systems operation ranging from transmission and distribution optimization for optimizing the routing and scheduling of electricity through the power grid to resilience optimization to recover from various types of disruptions, from equipment failures to natural disasters and cyber attacks. 

Quantum optimization tools are expected to achieve super-polynomial advantage for solving combinatorial optimization problems. Hence, quantum optimization tools have been studied in literature for solving power system problems. One example is presented in Ref.\  \cite{colucci2023power} where they studied a power grid optimization problem on D-Wave quantum annealer's constrained and binary quadratic model solvers. To this end, they needed to map the problem onto a QUBO problem. However, the problem involves both equality and inequality constraints. They introduced a method to translate inequality constraints with real-valued coefficients into approximate penalty functions in order to reduce the number of required slack variables. They also partitioned the problem into smaller graph partitions and demonstrated that hybrid quantum-classical methods outperformed classical methods in terms of solution quality for the problem studied.

In this work, we identified our optimization problem of interest as solving weighted MaxCut problems and left other optimization problems in power systems as future work. Determining maximum power exchange sections is one of the most interesting real-life applications of weighted MaxCut problems. The maximum power section of a power grid system is an important index for identifying a system's power delivery capability \cite{jing2023data}, and in the case of a microgrid, it determines the dependence of each microgrid on other microgrids~\cite{jing2022quantum}. 

The maximum power sections problem has been studied in the literature using QAOA. The authors of Ref.\ \cite{jing2023data} studied the IEEE 24-bus system and proposed a data-driven QAOA that yielded results comparable to those by the Goemans-Williamson algorithm. Similarly, in Ref.\ \cite{jing2022dynamics}, they performed a dynamic analysis of a microgrids test system with 20 buses  that included 7 EV charging stations using QAOA. By changing the power demands of these EV charging stations, they studied the maximum power sections. In Ref.\ \cite{jing2022quantum}, they switched their focus to a disturbance analysis of microgrids using QAOA. Inspired by this research, we studied maximum power section identification in IEEE 9- and 14-bus systems.

% To formulate it as a  QUBO problem, the authors of Ref.\ \cite{wang2022quantum} utilized Lagrange relaxation for inequality constraints and alternating direction method of multipliers to transform the integer quadratic programming problem at hand.
% Power grid optimization as a QUBO problem was also studied in Ref.\  \cite{colucci2023power}.

% In \cite{fernandez2021community} work explores its application in Graph Partitioning using electrical modularity. They benchmarked several quantum annealing and hybrid methods on IEEE well-known test cases. 

% Other use cases that can be solved using the MaxCut problem include dynamic analysis of microgrids with EV charging stations \cite{jing2022dynamics} and finding maximum sections of power and data delivery which is essential for monitoring, operation, and control of power grids \cite{jing2023data}. Ref.~\cite{jing2022quantum} also studies finding maximum power exchange sections of networked microgrids using QAOA. They provide graph topology for networked microgrids. The weights of the graph are obtained using power flow equations. However, it is not clear which dataset was utilized for the power flow equations.

% In \cite{Osaba_2023}, they have bin packing dataset and \cite{10422581} is the paper that uses this dataset.

%\subsection{Power Grids}
Power grid optimization systems can be solved by mapping the transmission system to a weighted graph and solving the weighted MaxCut problem \cite{Tong_2021}. The edge weights of the corresponding graph are given by the line impedances as 
\be w_{ij}=\begin{cases}
    \frac{1}{\sqrt{R_{ij}^2+X_{ij}^2}} \ , & \mathrm{nodes}  \ i \ \mathrm{and} \ j \ \mathrm{connected} \\ 
    0 \ , & \mathrm{otherwise}\\ 
\end{cases} \ee 
where $R_{ij}$ is the line resistance and $X_{ij}$ is the line reactance. These are parameters that are provided in the IEEE test case datasets \cite{matpower-data}.  

% \subsection{Charging Stations}

\section{Solving MaxCut Problem on Analog Quantum Computers}\label{sec:2methods}
Analogue Hamiltonian simulation is an indispensable tool  for the investigation of the features and behavior of a given quantum system. It is implemented by manipulating the parameters of another quantum system that bears similarities to the one being studied and minimizing the cost function that depends on these parameters \cite{AWSGitHubRepo}. The methods studied in the literature include, but are not limited to, using stochastic optimizers, and Bayesian optimization \cite{finvzgar2023designing}. We identified two methods of solving the MaxCut problem, i.e., finding the maximum cut value of a graph of interest, on analog quantum computers: by (a) pulse shaping using machine learning, and (b) fine tuning of the local detuning parameters. In what follows, we discuss the details of these two methods. 

%\subsection{Problem Definitions}
The MaxCut problem is defined as follows. Given a graph $G=(\mathcal{V}, \mathcal{E})$, the graph is cut into two disjoint sets $\tilde{\mathcal{V}}$ and $\mathcal{V}'$ such that $\tilde{\mathcal{V}} \subset \mathcal{V}$ and $\mathcal{V}'=\mathcal{V}\backslash \tilde{\mathcal{V}}$. Then the subset of edges that correspond to these set of vertices are $\tilde{\mathcal{E}}$ and $\mathcal{E}'$. Similarly, the subset of edges are connected to the vertices as follows.
\begin{equation}
    \tilde{\mathcal{E}}=\{ \{i,j\} \in \mathcal{E} | i \in \tilde{\mathcal{V}} \oplus j \in \tilde{\mathcal{V}}\} \ ,
\end{equation}
where $\oplus$ is the logic operator exclusive or. Then the Max-Cut problem can be defined as a combinatorial optimization problem that aims to split the graph into two disjoint subsets such that the number of edges spanning the two subsets is maximized. This problem can be generalized to include weights $w_{ij}$ on the edges ($(i,j) \in \mathcal{E}$) and represented as a QUBO problem
with the cost function defined as
\begin{eqnarray}
    \bm{\mathcal{C}} &=&  \sum_{(i,j) \in \mathcal{E}} w_{ij} Z_i Z_j \ ,  \label{eq:cost}
\end{eqnarray}
for given weights $w_{ij}$.
We will minimize this cost function by utilizing the evolution of a two-dimensional array of Rydberg atoms which is governed by the time-dependent Hamiltonian
\begin{eqnarray}
    H(t) &=& \frac{\Delta_{\text{global}}}{2}\sum_{i=1}^n Z_i+\frac{\Omega}{2}\sum_{i=1}^n\left[X_i \cos \phi-Y_i \sin \phi\right] \nonumber\\
    &&+\frac{C_6}{4}\sum_{i<j}\frac{(Z_i-\mathbb{I})(Z_j-\mathbb{I})}{\left[(x_i-x_j)^2+(y_i-y_j)^2\right]^3} \ , \label{eq:hami}
\end{eqnarray}
where $\Delta_{\text{global}}, \Omega, \phi$ are, in general, time-dependent parameters whose time evolution is controllable.
\subsection{Pulse Shaping}
\label{sec:pulse}
\begin{figure}[ht!]
    \centering
    \includegraphics[width=0.45\textwidth]{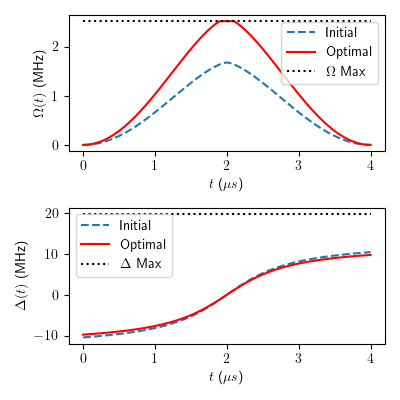}
    \caption{Amplitude ($\Omega$) and detuning ($\Delta$) as functions of time with default parameters (blue dashed line) and optimized parameters (red solid line) for the IEEE 9-bus problem. The black dotted line denotes the maximal amplitude and detuning values currently allowed on QuEra's Aquila processor. }
    \label{fig:pulse}
\end{figure}

The pulse shaping procedure finds the ground state by first finding an optimal register layout, then optimizing the amplitude and detuning pulses using the best layout. An adiabatic-type Ansatz \cite{aws-notebook-2} is chosen for the amplitude and detuning parameters as functions of time, where 
\bea \Omega(t,\bm{p}) &=& p_0 \left(1-\left[1-\sin\left(\frac{\pi  t}{t_{\mathrm{max}}}\right)^2\right]^{\frac{p_1}{2}}\right) \ , \nonumber\\ \Delta_{\text{global}} (t,\bm{p}) &=& \frac{2}{\pi} p_2 \tan^{-1}\left(p_1  \left[t-\frac{t_{\mathrm{max}}}{2}\right]\right) \ , \eea
and $\bm{p} = (p_0,p_1,p_2)$ is a vector of hyperparameters for the optimization. We set $t_{\mathrm{max}} = 4~\mu$s. The amplitude $\Omega(t,\bm{p})$ is the hopping parameter, and must start and end at $0$ MHz on the Aquila hardware. The detuning $\Delta(t,\bm{p})$ is the onsite operator, and is turned on gradually to slowly evolve the system from $\ket{0}^{\otimes N}$ to the MaxCut ground state. These functions before and after optimization for the IEEE 9-bus problem are shown in Fig. \ref{fig:pulse}. Note that there is also a maximum range of the detuning and amplitude, where $0\leq \Omega\leq 2.51$ MHz, and $|\Delta|\leq 19.89$ MHz, and these bounds are enforced in the optimization. 

After choosing a set of default hyperparameters, $N_R$ random layouts of graph vertex positions are chosen and the initial cost function is measured after evolving with the adiabatic evolution using the default parameters. Then, the layout that produces the lowest cost function is chosen for the pulse shape optimization step. The random layouts are found using an adaptation of the Fruchterman-Reingold (FR) layout \cite{FR_1991}. 

The graph embedding into the two-dimensional plane heavily influences the ability to find the ground state. Since the quantum Hamiltonian corresponding to the MaxCut problem is minimized when the atoms with edges between them are in opposite basis states, the graph embedding should place atoms with edges between them close together (based on edge weight) and atoms without edges between them far apart. The FR force directed algorithm \cite{FR_1991} finds the optimal node placement by relaxing a classical system with a repulsive force between nodes, $f_r$, and an attractive force between nodes with edges, $f_a$, given, respectively, by 
\be f^r_{ij}(d_{ij})=-k^2/d_{ij} \ , \ f^a_{ij}(d_{ij})=d_{ij}^2/k  \ , \ee 
where $d_{ij}$ is the distance between nodes $i$ and $j$, and $k$ is the optimal distance between nodes, with $k\propto \sqrt{\mathrm{Area}/\mathrm{\# vertices}}$. 
In this work, we introduce a variant of the FR algorithm that mimics the $r^{-6}$ scaling of the Rydberg interaction and also considers graph structures, where 
\be f^r_{ij}(d_{ij})=-k_{ij}^6/d_{ij}^6 \ , \ f^a_{ij}(d_{ij})=d_{ij}^2/k_{ij} \ , \ee 
Here, $k_{ij}$ is now local and dependent on the vertices $i$ and $j$, and can be determined by graph structure parameters such as connectivity, diameter, and cliques. We obtained the best results by determining $k_{ij}$ at each pair $(i,j)$ based on whether the pair $(i,j)$ is part of a 3-node clique, or triangle. We used the form 
\be  k_{ij}\propto \left(\frac{1}{N_\Delta}\sum_\Delta \{1 | i,j \in \Delta \} \right)^{\rho} \ ,\ee 
with the exponent $\rho$ a hyperparameter that was optimized. 
 
The pulse optimization was performed by optimizing the hyperparameters $\bm{p}$ using Pennylane's Nadam optimizer \cite{aws-notebook}. The optimization steps were simulated using Pennylane's Jax interface, and the optimized pulse shapes were then run on QuEra's Aquila device with 1000 shots \cite{aws-notebook-2}. 

\begin{figure}[ht!]
    \centering
      \subfloat[]{\includegraphics[width=0.45\textwidth]{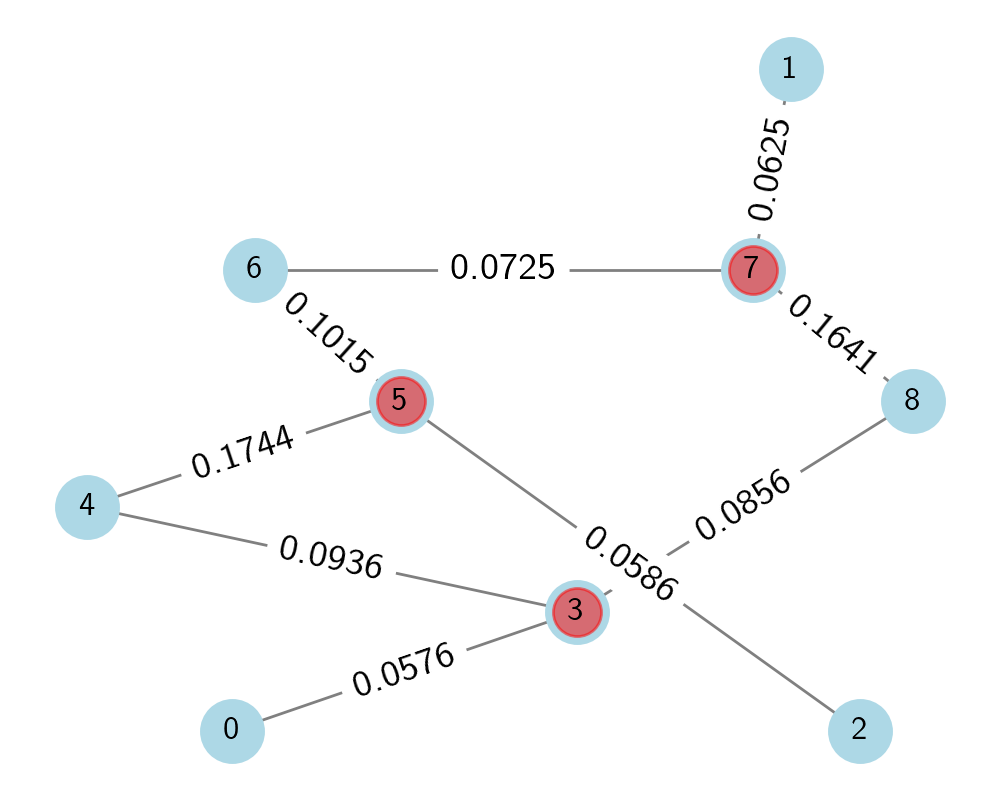}}\\
      \subfloat[]{\includegraphics[width=0.45\textwidth]{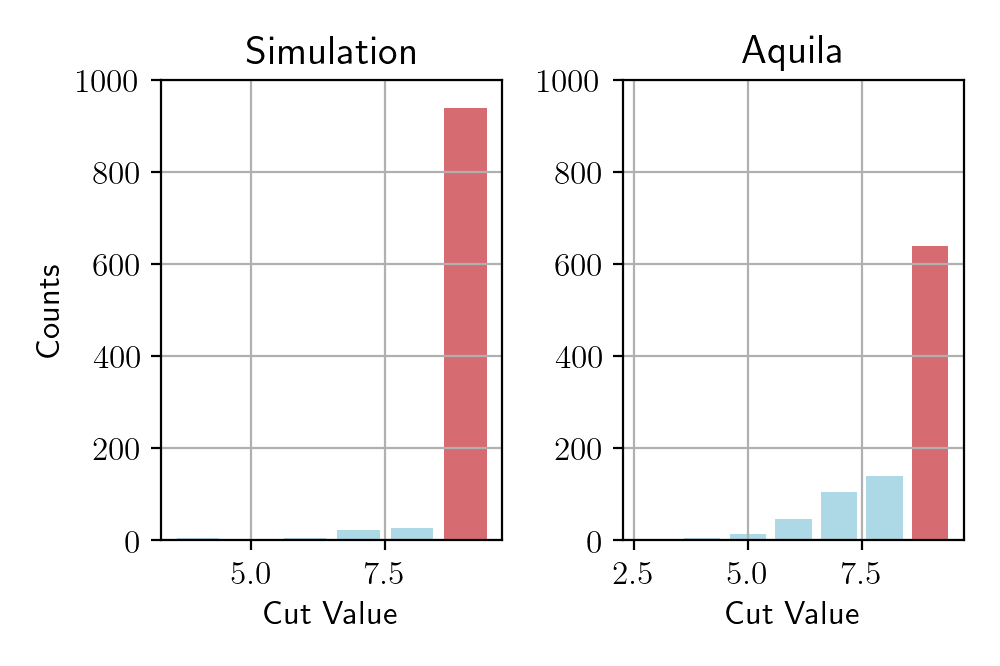}}
    \caption{(a) IEEE 9-bus dataset graph where each node represents a bus, weights correspond to the impedance calculated from dataset using $Z=\sqrt{X^2+R^2}$. The Max-Cut solution for the dataset presented in left panel, where the red and blue nodes correspond to different sets (b) Simulated and experimental histograms for the IEEE 9-bus problem after performing the optimization procedures. %(Here we can add the 14 bus problem result below).  
    } 
    \label{fig:IEEE9}
\end{figure}

For the IEEE dataset maximum power section optimization problems, the problem can be mapped to a weighted MaxCut problem. 
%We considered the IEEE 9-, 14-, and 20- bus problems. 
The weighted graph corresponding to the IEEE 9-bus problem is shown in Fig.\ \ref{fig:IEEE9}(a). The solution is given by the node color, where the red and blue nodes correspond to different sets. The experiment is performed using the simulator as well as the Aquila hardware with $N_R=50$, and the histograms are shown in Fig.\ \ref{fig:IEEE9}(b). We obtained the probability of measuring the maximum cut (overlap with the ground state), $P(GS)=0.939$ for the noiseless simulation, and $P(GS)=0.639$ using quantum hardware. Thus, despite degradation of performance, it was still possible to find the solution to the problem using the Aquila quantum hardware. 

\begin{figure}[ht!]
    \centering
    \subfloat[]{\includegraphics[width=0.45\textwidth]{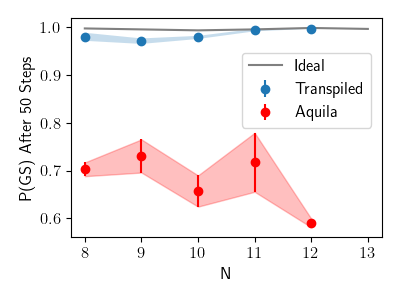} }\\
    \subfloat[]{\includegraphics[width=0.45\textwidth]{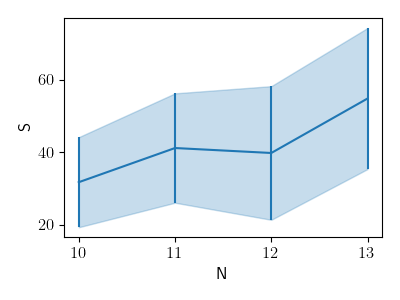}} 
    \caption{(a) Probability of measuring the ground state after $N_R=50$ steps for $8\leq N\leq 12$ vertex graphs, where the grey solid line represents the ideal result, the blue dots represent the result after the ideal evolution is fit within the hardware parameters, and the red dots represent the experimental results measured  on the Aquila processor. The bars denote standard error from averaging over random Erdos-R\'enyi graphs. (b) Step to solution $S$ as defined in Eq.\ \eqref{eq:step} for $10\leq N\leq 13$ vertex graphs. The bars denote standard error from averaging over random Erdos-R\'enyi graphs.  }
    \label{fig:ergraphs}
\end{figure}

In order to study the scaling of the performance with graph size, we also tested our method on Erdos-R\'enyi graphs with $p=0.25$ and weights selected from the uniform distribution $[0,0.25]$, which we found to be similar in complexity to the graphs corresponding to the IEEE bus problems. We chose $N_R=50$ optimizer steps and computed the average probability of measuring the maximum cut state $P(GS)$ for graphs with $8\leq N\leq 12$ vertices. Results are given for the ideal case and experimental results with standard error bars, shown in Fig.\ \ref{fig:ergraphs}(a). We find the probability of measuring the solution > 95\% for the noiseless simulations, and the average probability of measuring the solution is 60\% on Aquila quantum hardware. 

We also computed the average step to solution $S$, given as 
\be S=\frac{\log(0.01)}{\log(1-\bar{P}(GS))} \ , \label{eq:step}\ee 
which gives the average number of register layout optimization steps required to obtain the solution with 99\% certainty, shown in Fig.\ \ref{fig:ergraphs}(b). For the graphs tested in the range $10\leq N\leq 13$, we see an approximately linear scaling in the amount of register layout steps required to find the solution. 
%For solving MaxCut, \cite{coelho2022efficient} uses up to 200 ``register optimization" steps where they try different placements of the atoms until ground state convergence is reached. This will be resource-intensive. Below, a maximum of 5 register optimization steps are used for $N$ vertex Erdos-Renyi graphs with p=0.5. 30 graphs are tested for each system size, and the average P(GS) is calculated (blue). The green line gives the probability that convergence to the ground state with P(GS)$>0.95$ will be obtained. 

%We can obtain some improvement by doing 100 register optimization (RO) steps, where we only use 1 run on Aquila to determine the quality of the layout. Then, the best layout is taken and adiabatic procedure is used to find optimized pulse parameters. The result for average P(GS) is shown as the purple line. 

%I am working on parameterizing the register layout so that we can optimize over some set of parameters and use less steps. MaxCut is said to be very dependent on the presence of triangles and odd cliques in the graphs, so one idea is to write layout algorithm that treats triangles specifically. 

%Studying finding maximum power sections using MaxCut for IEEE bus dataset (see Fig.~\ref{fig:IEEE9}). \textcolor{red}{Is this sentence misplaced?}

%\begin{figure}
%    \centering
%    \includegraphics[width=0.7\textwidth]%{QuEra_Hardware_1.png}
%    \caption{Experimental and simulated results for %the IEEE 9 dataset graph Max-Cut problem presented in Fig.~\ref{fig:IEEE9}}
%    \label{fig:IEEE9-solution}
%\end{figure}

\subsection{Local Detuning}
Our second approach to the MaxCut problem was by using the QAOA. In the analog quantum computer of neutral atoms, we define the cost and mixer time evolution operators by fine-tuning the detuning, the Rabi frequency and the laser phase. The solution to the problem can be improved by assuming access to local detuning. In this case, we can re-write the Hamiltonian, Eq.\ \eqref{eq:hami},  as
\begin{eqnarray}
    H &=& \sum_{i=1}^n \left({\frac{\Delta_i}{2} -\sum_{j\neq i} w_{ij}}\right) Z_i +\sum_{i<j} w_{ij} Z_i Z_j \nonumber\\
    && +\frac{\Omega(t)}{2}\sum_{i=1}^n\left[X_i \cos \phi (t) -Y_i \sin \phi (t)\right] \ , \label{eq:hami2}
\end{eqnarray}
with distance-dependent weights given by 
\begin{equation}
    w_{ij}= \frac{C_6}{4\left[(x_i-x_j)^2+(y_i-y_j)^2\right]^3} \label{eq:weight} \ .
\end{equation}
Notice that by setting the local detuning parameters to $\Delta_i = 2\sum_{j\neq i} w_{ij}$, we can completely eliminate the linear $Z$ terms from Eq.\ \eqref{eq:hami2}. These values of local detuning parameters will be assumed in the following discussion.

To apply the QAOA, we prepared the atoms in the ground state $\ket{0}^{\otimes n}$ and let the system evolve for time $t_0=\frac{\pi}{2 \Omega_0}$ while fixing $\phi(t)=-\frac{\pi}{2}$ and $\Omega(t) =\Omega_0$. With these choices, the system's Hamiltonian became
\begin{eqnarray}
    H_0 &=&  \frac{\Omega_0}{2}\sum_{i=1}^n Y_i  +\sum_{i<j} w_{ij} Z_i Z_j  \label{eq:hami3}\ .
\end{eqnarray}
After time $t=t_0$, the state of the neutral atoms evolved from the product state $\ket{0}^{\otimes n}$, to the uniform superposition $\ket{+}^\otimes$, to a good approximation. This is because the amplitude was chosen to be $\Omega_0 = 5\pi$ MHz, whereas the maximum weight was $w= 0.45$ MHz, corresponding to minimum distance between two vertices (nearest neighbors)  $\alpha = 12~\mu$m, therefore $w_{ij} \ll \Omega_0$. 

Next, we introduce the cost Hamiltonian by setting $\Omega(t)=0$
\begin{eqnarray}
    H_c &=&  \sum_{i<j} w_{ij} Z_i Z_j  \ ,\label{eq:hami4}
\end{eqnarray}
and the mixer Hamiltonian by choosing $\Omega(t)=2$ MHz (which was the best performing value for smaller graphs) and $\phi (t)=0$
\begin{eqnarray}
    H_b &=& \sum_{i=1}^n X_i + \sum_{i<j} w_{ij} Z_i Z_j  \ .\label{eq:hami5}
\end{eqnarray}
A single QAOA layer is implemented by time evolution under the cost Hamiltonian for  $t_1 = \gamma_1$  and mixer Hamiltonian for time $t_2= \beta_1$. After applying $p$ layers we obtain the state
\begin{equation}\label{eq:14}
    \ket{\psi_p}=e^{-iH_b \beta_{p}} e^{-iH_c \gamma_{p}}  \cdots e^{-iH_b \beta_1} e^{-iH_c \gamma_1} \ket{\psi_0} \ .
\end{equation}
At the end, we minimize the value of the cost function,
\begin{equation}
    \mathcal{C}_p = \braket{\psi_p | \bm{\mathcal{C}} | \psi_p} \ ,\label{eq:cost_p}
\end{equation}
where the cost function is defined in Eq.\ \eqref{eq:cost}.
It is important to understand that there is a difference between the cost function (Eq.\ \eqref{eq:cost}) we are trying to minimize and a similar set of terms in the Hamiltonian \eqref{eq:hami2} we are utilizing. The latter includes all the interactions of the $n $ body quantum system which fall off rapidly according to the van der Waals law, $\sim r^{-{6}}$. The sum in the cost function has only terms that correspond to an edge on the graph we are interested in. Because of this, the performance of QAOA depends significantly on the position of each atom and how well they represent the graph, including weights, on which the cost function is defined.

Instead of trying to find the optimal position of atoms based on the FR algorithm, as indicated above, we considered the placement of the atoms at fixed lattice points. 
Using this technique, we solved the MaxCut problem for random graphs with a number of vertices $\in[10,13]$ generated from square and honeycomb lattices, and for the IEEE 9-bus dataset. Specifically, for the random graphs, we considered two graphs for each lattice geometry.

\begin{figure}[ht!]
    \centering
    \includegraphics[width=0.45\textwidth]{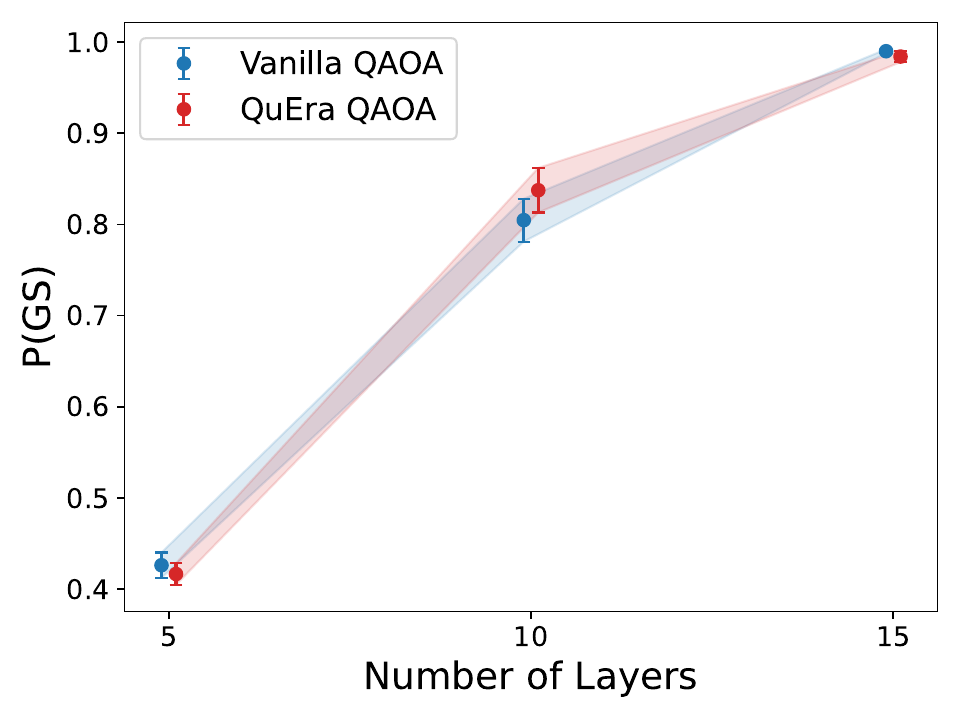}
    \caption{Probability of measuring the ground state for the IEEE 9-bus graph. For both simulations (Vanilla QAOA and QuEra) we calculated the probability by applying 5, 10 and 15 layers. In each case, we started with 1000 optimization steps and increased then up to 2000.}
    \label{fig:IEEE9_loc}
\end{figure}

\begin{figure}[ht!]
    \centering
    \includegraphics[width=0.45\textwidth]{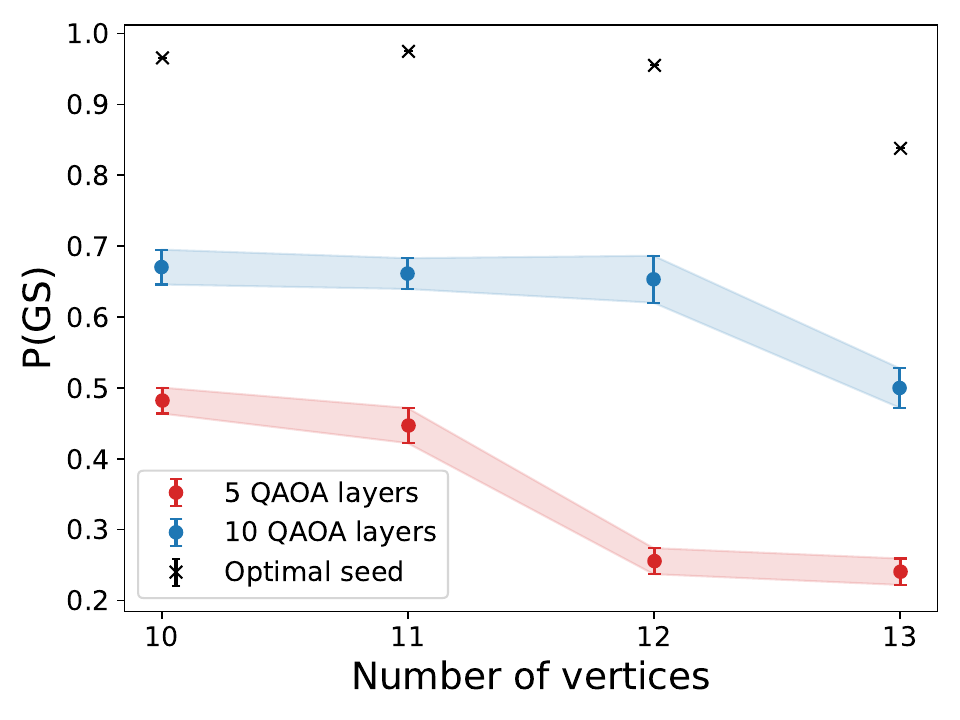}
    \caption{Probability of measuring the ground state for $10<N<13$ vertex graph's by simulating QuEra's hardware. For $5$ layers optimized the cost function up to $1000$ steps using the Adam Optimizer, whereas for $10$ layers we considered up to $2000$ steps. }
    \label{fig:V10_15_1_loc}
\end{figure}

For numerical results, we calculated the cost function following the QAOA protocol and then used the Adam Optimizer to find the optimal values of $\beta_i$ and $\gamma_i$ in \eqref{eq:14}. For each graph, we considered 20 different initial seeds on these parameters and up to 2000 optimization steps. For the IEEE 9-bus graph, we find an average probability of measuring the ground state (which corresponds to the MaxCut) equal to $P(GS)=0.984$. We also evaluated the average probability to find the MaxCut for the vanilla QAOA \cite{farhi2014quantum}, in which there is no distance dependence and found $P(GS)=0.99$. These results are illustrated in Fig.\ \ref{fig:IEEE9_loc}. Finally, for the random graphs on the lattice found on average probability of measuring the ground state that decrease from $P(GS)=0.67$ to $P(GS)=0.5$ as we increase the number of vertices. However, on larger graphs these results depend stronger on the initial random seed. Because of this we have outliers where the optimization fails to converge for specific seeds. It is worth mentioning that in the best case scenario, there are initial seeds that give probability which decreases from $P(GS)=0.956$ to $P(GS)=0.838$. These results are summarized in Fig.\ \ref{fig:V10_15_1_loc}.

\begin{figure}[ht!]
    \centering
    \subfloat[]{\includegraphics[width=0.45\textwidth]{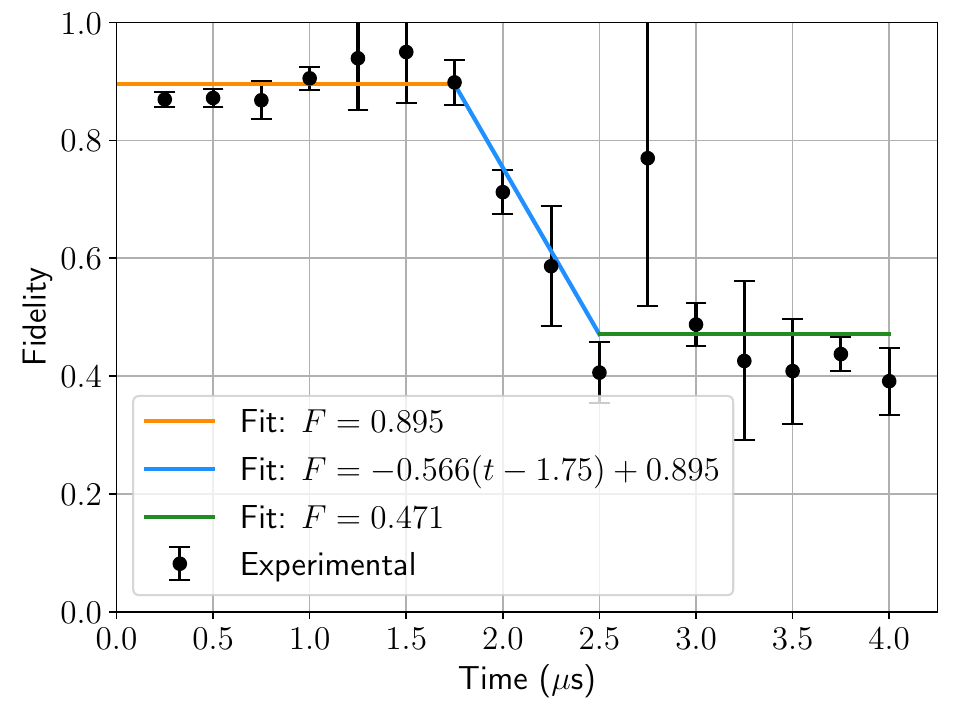}}\\
\subfloat[]{\includegraphics[width=0.45\textwidth]{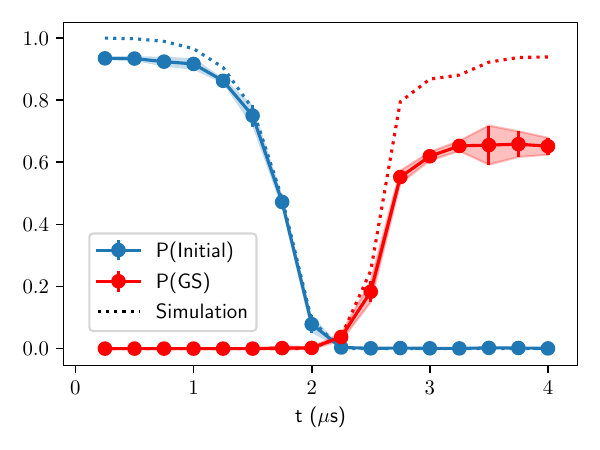}}
    \caption{(a) Fidelity values calculated in 16 time cycles where each time cycle is 0.25 $\mu$s. The experiments were conducted three times to obtain the error bars.  (b) Probability of measuring the initial state $\ket{0}^{\otimes N}$ (blue points) and probability of measuring the ground state (red points) as a function of time, as measured on hardware. The dotted lines correspond to the values measured in simulation. Error bars represent $3\sigma$ at each time step. 
}
    \label{fig:Fidelity}
\end{figure}

\section{Fidelity estimation on QuEra}\label{sec:benchmarking}
In this section, we present the fidelity estimation of the graph states that are studied in Section~\ref{sec:usecase}. Fidelity can be utilized as an important benchmark for studying quantum systems and can be defined as
\begin{equation}
    F=\braket{\psi|\rho_{\text{exp}}|\psi}~,
\end{equation}
where $\rho_{\text{exp}}$ is the experimental mixed state. On digital quantum computers, there is more than one way of computing the fidelity, i.e., using quantum state tomography, quantum process tomography, randomized benchmarking \cite{PhysRevA.77.012307}, or cross entropy benchmarking \cite{arute2019quantum}. On the other hand, there are also various protocols developed for fidelity estimation on analogue quantum computers, i.e., quantum simulators \cite{choi2023preparing,shaw2023benchmarking,mark2023benchmarking}.

We conducted a fidelity estimation using the following equation from Ref.~\cite{choi2023preparing}:
\begin{equation}
    F=2\frac{\sum_z p_0(z) p(z)}{\sum_z p_0^2(z)}-1~,
\end{equation}
where $p_0(z)$ is the theoretical and $p(z)$  the experimental probability values for each bit string. For experimental probability values we followed the protocol in Ref.\  \cite{mark2023benchmarking} which can be summarized as follows: 

{\bf{Experiment:}}
\begin{enumerate}
    \item Prepare an initial step $\rho_0$, which approximates a pure state $\ket {\psi_0}\bra{\psi_0}$.
    \item Evolve the system under its natural Hamiltonian $H$ for a time $t$.
    \item Measure the evolved state $\rho(t)$ in a natural basis, obtaining configurations $\{z_1, \dots, z_M\}$.
\end{enumerate}

In order to obtain the experimental probability values, we prepared the IEEE 9-bus dataset graph in the initial state $|\psi_0\rangle=|0\rangle^{\otimes N}$, as seen in Fig.~\ref{fig:IEEE9}, and time evolved it in 16 time cycles where each time cycle was $t_{\text{cycle}}=0.25~\mu$s. We measured the time evolved state in the computational basis at each time cycle. We used $n_{\text{shots}}=1000$ for our experiments run on QuEra's Aquila neutral atom quantum computer. On the other hand, for theoretical probability values, $p_0(z)$, we ran a noiseless quantum simulator with $n_{\text{shots}}=100,000$. We obtained the fidelity estimation values depicted in Fig.~\ref{fig:Fidelity}(a). 
% \begin{figure}
%     \centering
%     \includegraphics[width=0.45\textwidth]{FidelityTimeChoiEq.png}
%     \caption{Fidelity values calculated in 16 time cycles where each time cycle is 0.25$\mu$s.}
%     \label{fig:Fidelity}
% \end{figure}
Evidently, the fidelity remains relatively constant in the first $\approx 1.75$ $\mu$s, then decays and settles into another nearly constant value after $t\approx 2.5~\mu$s. This matches the behavior of the detuning and amplitude (see Fig.\ \ref{fig:pulse}) as a function of time: for the first 2 $\mu$s, the detuning has a negative value and the amplitude increases slowly whereas for $t > 2~\mu$s, the detuning has positive value and the amplitude decreases down to zero at $t= 4~\mu$s. Correspondingly, the probability of overlap with the initial and ground states, shown in Fig.\ \ref{fig:Fidelity}(b), changes around $t = 2~\mu$s. A fit of the experimental data on the fidelity (Fig.\ \ref{fig:Fidelity}(a)) reveals that for $t < 2~\mu$s, the fidelity stays fairly constant at 89.5\%. During this time the system remains approximately in the initial state which is not entangled. As it undergoes entanglement, the fidelity decreases down to 47.1\% and remains fairly constant for $t > 2~\mu$s. During this time, the system has large overlap with the ground state which is also not entangled. Thus, we observe a degradation in fidelity during the evolution of the system that involves an increase in entanglement. This result is an important benchmark of the neutral atom hardware. It should be pointed out that despite degradation, we still attain our goal, as we obtain the solution to the MaxCut problem with 60\% probability.

\section{Conclusion}\label{sec:conclusion} 

In this work, we investigated the performance of two methods we introduced for solving power grid optimization problems mapped onto weighted MaxCut problems using neutral atom quantum hardware. 

In the first method, we used a Rydberg interaction inspired graph embedding algorithm and Pennylane's Jax machine learning interface to optimize the pulse shape of an adiabatic Ansatz. We found that with under 100 embedding steps, the adiabatic algorithm was able to find the ground state with over 95\% overlap. The optimized pulse shapes were also tested on QuEra's Aquila processor, and we found an average 60\% overlap with the ground state experimentally. Although far from ideal, the performance of the QuEra processor was adequate to yield the solution to the problem at hand.

In the second method, we applied the QAOA on graphs which can be embedded into square or honeycomb lattices. Using the Adam Optimizer for the classical optimization and up to 10 QAOA layers we were able to find the MaxCut on average with probability ranging from 67\% (for 10 vertices) to 50\% (for 13 vertices). 

We also measured the experimental fidelity of the adiabatic evolution to the ground state for the IEEE 9-bus problem on the Aquila processor to be utilized as a benchmark.

With the improvements in neutral atom hardware, it is important to benchmark the algorithmic performance in order to compare with other quantum hardware. Schemes have been devised for quantum reservoir computing with neutral atoms \cite{PRXQuantum.3.030325}, using the native Hamiltonian for computation in a quantum recurrent neural network. Utilizing neutral atom hardware is a promising avenue for reservoir computing because the Rydberg interaction allows for a high level of entanglement, which in turn allows for more complex reservoir dynamics and a longer short-term memory \cite{PhysRevA.108.052427}. There have also been proposals for the quantum kernel method benchmarking on neutral atom quantum computers \cite{PhysRevA.104.032416}, which involves measuring the similarity between graph-structured data.

Since our optimization was performed on a simulator instead of the Aquila hardware, it is important to study the performance when the optimization is performed on actual hardware. Bayesian optimization  has been studied in optimizing pulse shapes in neutral atom hardware for the MIS problem \cite{finzgar_2023_designing}, so it would be interesting to compare the performance of Bayesian optimization with the gradient descent NAdam algorithm on MaxCut problems. Moreover, it is important to study error mitigation techniques specialized for analog processors, such as dual-state purification \cite{shingu2022quantum}. 

Additionally, the emergence of local detuning on analog neutral atoms systems increases the possibilities for quantum algorithms on neutral atom quantum computers. Local detuning has been shown to allow for single and two-qubit digital gates, SWAP networks, and analog variational quantum eigensolver (VQE) algorithms \cite{chevallier2024variational}. Using digital gates, there have also been proposals for quantum machine learning on neutral atom hardware, where the trainable Ansatz uses both digital and analog evolution, and numerical experiments are performed on handwritten digit classification and quantum phase boundary learning. It was found that the digital-analog schemes require shorter depth and are more robust to noise \cite{lu2024digitalanalog}. It is important to apply these tools to real-life applications, such as power grid optimization and placement of EV charging stations that were discussed here, in search of quantum utility.
%Quantum annealing with error mitigation \cite{shingu2022quantum}. 
 
%Exploring quantumness in quantum reservoir computing \cite{PhysRevA.108.052427}.

%Quantum Reservoir Computing Using Arrays of Rydberg Atoms \cite{PRXQuantum.3.030325}.

%Digital-analog quantum learning on Rydberg atom arrays \cite{lu2024digitalanalog}. 

%Schedule optimization of neutral atom quantum computers using Bayesian optimization  \cite{aws-blog}.

%In \cite{PhysRevA.104.032416} they studied machine learning, specifically quantum kernel method, with neutral atom quantum computers. They study measuring the similarity between graph-structured data.

%\newpage
%\newpage
%\setcounter{page}{1}

\acknowledgments
Research supported by the National Science Foundation under award DGE-2152168, the Department of Energy under award DE-SC0024325, and the DARPA ONISQ Program under award W911NF-20-2-0051. 
A portion of the computation for this work was performed on the University of Tennessee Infrastructure for Scientific Applications and Advanced Computing (ISAAC) computational resources. The Aquila processor was accessed through Amazon Braket, and we acknowledge support from the AWS Cloud Credit for Research program. 

KYA was supported by the Department of Energy under award DE-SC0024451 and MITRE's Quantum Horizon Program. \copyright 2024 The MITRE Corporation. ALL RIGHTS RESERVED. 
Approved for public release. Distribution unlimited PR-24-00320-1.

%\section*{Appendix 3: Bibliography}

%\bibliographystyle{apsrev4-2} 
%\bibliography{bibliography}


\end{document}